\documentclass[aps,pra,twocolumn,superscriptaddress]{revtex4-1} 

\usepackage{amsmath}
\usepackage{amsfonts,amssymb,amsthm, bbm,braket}
\usepackage{graphicx}   
\usepackage{subfigure}  
\usepackage{amsbsy} 
\usepackage[bold]{hhtensor} 
\usepackage{natbib}
\usepackage{algpseudocode}
   
\usepackage{multirow}

\usepackage[normalem]{ulem}

\newcommand\Tr{\mathrm{Tr}}

\usepackage{hyperref} 

\hypersetup{
  colorlinks   = true, 
  urlcolor     = blue, 
  linkcolor    = blue, 
  citecolor   =  red 
}

\begin{document}

\title{Robust and efficient \emph{in situ} quantum control}

\author{Christopher Ferrie}
\email{csferrie@gmail.com}
\affiliation{
Center for Quantum Information and Control,
University of New Mexico,
Albuquerque, New Mexico, 87131-0001}
\affiliation{Centre for Engineered Quantum Systems, School of Physics, The University of Sydney, Sydney, NSW, Australia}

\author{Osama Moussa}
\email{omoussa@gmail.com}
\affiliation{
Institute for Quantum Computing
and 
Department of Physics and Astronomy, 
University of Waterloo, Waterloo, ON, N2L 3G1, Canada
}

\date{\today}
%
%
\begin{abstract}
Precision control of quantum systems is the driving force for both quantum technology and the probing of physics at the quantum and nano-scale.  We propose an implementation independent method for \emph{in situ} quantum control that leverages recent advances in the direct estimation of quantum gate fidelity.  Our algorithm takes account of the stochasticity of the problem and is suitable for closed-loop control and requires only a constant number of fidelity estimating experiments per iteration independent of the dimension of the control space.  It is efficient and robust to both statistical and technical noise.
\end{abstract}


\maketitle

\section{Introduction}

Precision quantum control enables practical goals such as quantum computation \cite{nielsen2010quantum} and quantum metrology \cite{caves1981quantum} which in turn can provide probes of fundamental physics such as  gravity wave detection \cite{aasi2013enhanced}.  To process the information in a quantum device requires enacting quantum logic gates with high fidelity.  State of the art methods calculate the solution to Schrodinger's equation in iterative gradient climbing algorithms \cite{khaneja2005optimal} to arrive at a control pulse that has high fidelity to some target evolution with respect to some physical model.  { Contrary to this approach, which we call \emph{ex situ} control, it would be convenient if the quantum device guided itself to a desired state---\emph{in situ} quantum control.  

Control of the \emph{in situ} type can be divided into two categories: those that use a fresh copy of the system with each measurement and those that use a single copy in a continuous measurement scenario.  The former was pioneered by Rabitz \emph{et al} in the context of state transfer using genetic algorithms \cite{judson1992teaching}, and is the type of \emph{in situ} control we consider.  The latter is often called \emph{quantum feedback control} and the interested reader is referred to \cite{WisemanMilburn2010}.  Both are also referred to as either \emph{feedback} or\emph{closed-loop} control \cite{brif_2010_control}.
}

Until recently, however, it was not known whether the fidelity to some target gate could be estimated efficiently from experiment.  Here we leverage recent advances in fidelity estimation \cite{emerson2007symmetrized, flammia2011direct, da2011practical, moussa2012practical, magesan2012efficient} to design an \emph{in situ} quantum control algorithm which can in principle efficiently and robustly find the optimal control sequence.  In particular, we show via numerical experiments that our \emph{in situ} algorithm converges in fidelity to the target unitary gate at rate given by $O(1/ N_{\rm tot})$ where $N_{\rm tot}$ is the total number of experiments performed.

In very broad strokes, the current paradigm for quantum control proceeds in four steps: (i) modeling, (ii) estimation, (iii) optimization and (iv) implementation and verification.  At each of these steps, many things can go wrong.  We briefly overview each step, recall what could go wrong and describe how \emph{in situ} control remedies the problem.

\begin{figure}[t]
\includegraphics[width=.75\columnwidth]{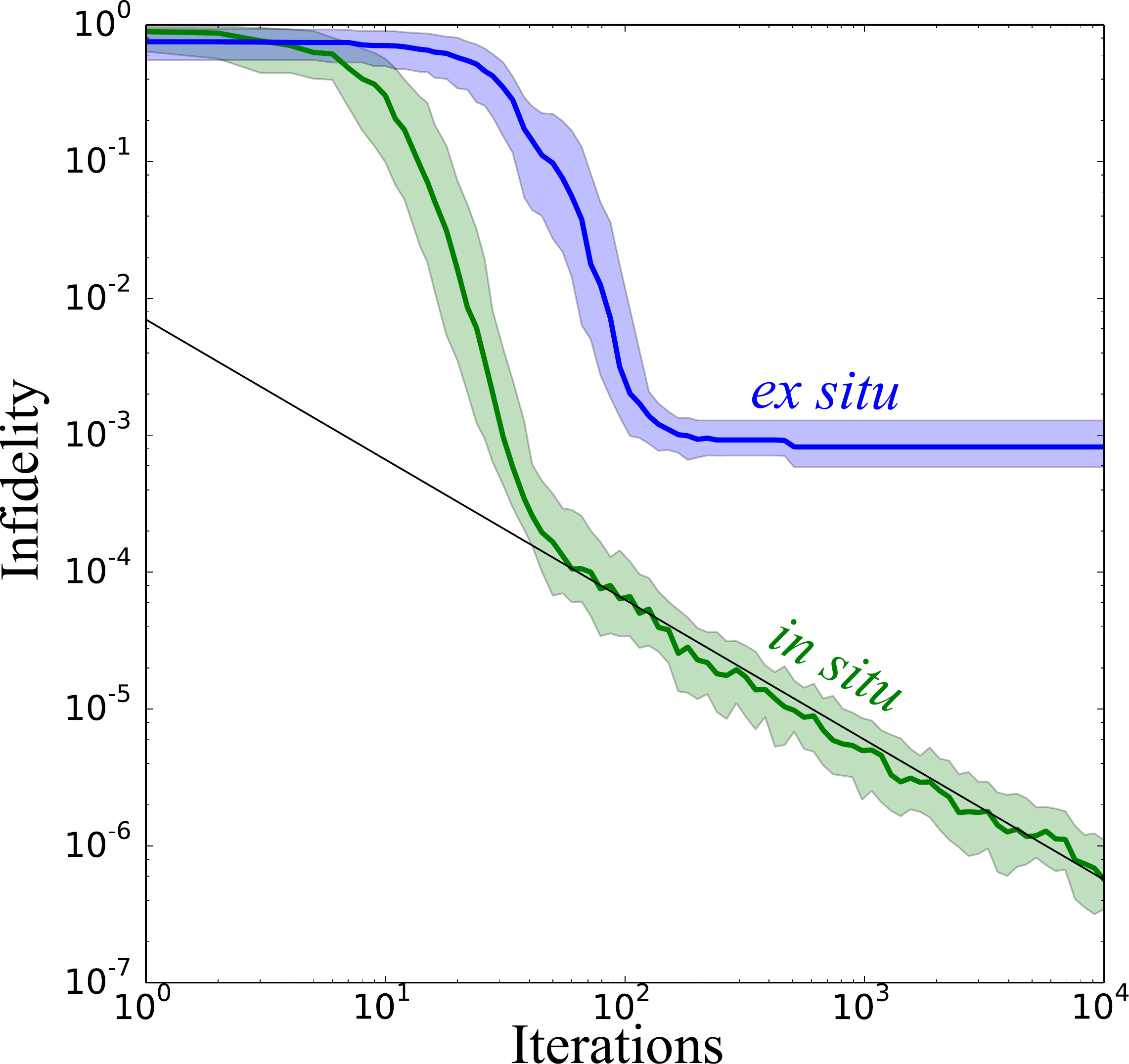}
\caption{\label{fig:first} (Color online)
(Color online) The median \emph{infidelity} (1-fidelity) of our \emph{in situ} algorithm and an \emph{ex situ} algorithm as a function of the number of iterations.  The problem is that of designing a random single qubit gate and is described in more detail in Section \ref{sec:examples}.  The \emph{in situ} algorithm was afforded only $10^3$ experiments per fidelity estimate while the \emph{ex situ} algorithm was given the ability to compute the fidelity to machine precision but saturates since the assumed Hamiltonian was incorrect by $||\triangle H || = 10^{-2} || H ||$.}
\end{figure}

The first task is to model the system; that is, to create a physical model of the dynamics experienced by the system.  Much can be learned from our current understanding of the physics involved. However, as physics is the art of approximation, the model will never be precisely correct.  It can be very good, but the precision in control will be limited by the precision in the model \cite{castro2012controlling}.  Alternatively, if the system itself is used, then the ``model'' used to design controls is perfect---it is an exact replica of the physics because it is the physics!  Thus, there are no limitations due to modeling errors for \emph{in situ} control.  This is illustrated in Fig.~\ref{fig:first}.

Given a model, the next task is to estimate the parameters of the model.  There are many ingenious ways to do this (from full process tomography \cite{weinstein2004quantum} to Hamiltonian learning using quantum resources \cite{wiebe2013hamiltonian}).  Although estimation of physical parameters is limited in precision due to statistical and technical errors, control algorithms can be made ``robust'' to these imperfections.  But there is no free lunch here either: a control strategy that is robust to a range of parameters is typically more taxing to find, and will not be strictly optimal for the ``true'' parameters \cite{egger2014optimized}. For \emph{in situ} control, we have noted already that a model is not required.  Therefore, the task of estimating the parameters of the model is not required and any problems incurred by this step are nonexistent.

Given a model and set of parameters, the task now is to find the optimal set of controls.  This is usually done through some iterative optimization algorithm performed on a classical computer \cite{khaneja2005optimal}.  For each candidate set of controls, a classical simulation is required.  It is thus inefficient in general to solve such an optimization problem.  Moreover, it is doubly inefficient since a classical simulation is used to propagate the state vector only to compute a single number.  While \emph{in situ} control trivially sidestepped the previous problems, here is where it provides a technical advantage.  Since the system itself will perform its own simulation, it may seem efficient.  However, the key difference between classical simulation (as used in control design) and quantum simulation is that a quantum simulator can only produce sample outcomes.  Thus, it may be the case that an inefficient number of experiments need to be run at each step in the optimization protocol.  The main result of this paper is to show this is not the case.  That is, our \emph{in situ} control strategy is efficient.

Finally, given the optimized set of controls, the final task is to implement those controls in an experiment and verify that the intended gate was produced.  High precision experiments are an impressive feat of physics and engineering---there are many imperfections in machined or fabricated components, stray fields and vibrations, noisy electronics and so on that need to be overcome.  Thus, it is not often the case that the optimal controls will reach the same value of the objective function in reality that they did in simulation. For \emph{in situ} control, once the optimal control has been found, we are done.  That is, the last experiment performed in our iterative optimization algorithm is the one sought after and verifies its own performance.

The imperfections due to the classical control devices are of systematic or stochastic origins. The systematic deviations (resulting from e.g. nonlinear or non-uniform response of amplifiers or resonators) can in principle be reversed/accounted for with pre-distortions of the optimized sequences, or using classical feedback, or even by inclusion in the modeling or optimization steps. The random distortions, or the \emph{technical noise}, plagues the implementation and verification step for \emph{ex situ} schemes, and countering this requires more data and averaging, which  scales quadratically with the desired precision. The technical noise challenges \emph{in situ} control in a different way---if the decision making of the optimization method is rigid and based solely on a deterministic view of the fidelity landscape, it is not hard to imagine how it would be limited by randomness in the fidelity evaluations. Thus, it is important for our scheme to allow for the stochastisity of the fidelity estimation.

Although in the standard paradigm the challenges in modeling, estimation and implementation are in principle avoidable in a perfect world, those in actual control finding are not.  That is, in general, classical optimization of the controls is inefficient.  For our scheme to provide a solution to this problem, we need to show that it is efficient.  It is illustrative to compare to two recent similar protocols, which use both classical and \emph{in situ} control: Ad-HOC \cite{egger2014adaptive} and ORBIT \cite{kelly2014optimal}. Both protocols rely on the Nelder-Mead (NM) optimization algorithm. Since these have such sexy names, we need to give our algorithm a name---we call it Adaptive Control via Randomized Optimization Nearly Yielding Maximization (ACRONYM).

ACRONYM can be divided into two distinct pieces: fidelity estimation and stochastic optimization.  Fidelity estimation can be achieved by various means.  For certain classes of gates, efficient algorithms exist \cite{flammia2011direct, da2011practical, moussa2012practical, magesan2012efficient} and have already been experimentally implemented~\cite{ryan2009randomized,moussa2012practical,chow2009randomized,gaebler2012randomized,steffen2012experimental}.  However, any fidelity estimation scheme requires many repeated experiments to reduce the noise due to statistical fluctuations.

In Ad-HOC, the noise is modeled by a depolarizing channel, and in ORBIT, the noise is some combination of technical experimental noise and statistical noise due to finite sampling.  The latter provides an ultimate lower bound on the achievable accuracy in Ad-HOC and ORBIT, whereas ACRONYM is not limited by statistical noise.  The reason, as observed in \cite{egger2014adaptive}, is due to the sensitivity of the NM algorithm to fluctuations in fidelity.  After describing ACRONYM, we demonstrate that it is efficient in this sense of requiring only a fixed number of experiments per iteration of the algorithm.  Broadly speaking, ACRONYM is implementation independent, requiring only that changes in the control space produce changes in the fidelity which can be estimated via experiment.  

The outline of the paper is as follows.  In Section \ref{sec:acronym} we introduce the problem and our control algorithm ACRONYM.  Section \ref{sec:examples} reports on numerical experiments benchmarking the performance of the algorithm.  Section \ref{sec:end} contains some further discussion and concludes the paper.

{
\section{ACRONYM: Adaptive Control via Randomized Optimization Nearly Yielding Maximization\label{sec:acronym}}
}
\subsection{Problem statement}

Generally, the task is to select a set of controls $\boldsymbol{c}\in \mathbb R^p$, where $p$ is the dimension of the control space, such that the implemented channel $\Lambda_{\boldsymbol{c}}$ is a close as possible to some target $U_{\rm T}$, assumed to be a unitary.  The ``closeness'' is measured by the channel fidelity $f(\boldsymbol{c}):=F(\Lambda_{\boldsymbol{c}}, U_{\rm T})$, where we have defined the object function $f$ to maximize.  When the implemented gate is also unitary---the case we consider---we have
\begin{equation}\label{eq:underlying}
F(U_{\vec{c}}, U_{\rm T}) = \frac{1}{d^2}|\Tr( U_{\rm T}^\dag U_{\vec{c}})|^2,
\end{equation}
where $d$ is the dimension of the quantum system.

A classical simulation can provide an exact numerical calculation of $f$, whereas a quantum simulation only provides a single datum---many from which we can estimate $f$.  Happily, there exists efficient protocols to estimate $f$---for example, randomized benchmarking \cite{magesan2012efficient}, direct fidelity estimation via Monte Carlo \cite{flammia2011direct, da2011practical}, or certification via twirling~\cite{moussa2012practical}.  Let us suppose that we are in some regime where we can consider a good estimate of the fidelity to be
\begin{equation}\label{f+noise}
\hat f(\boldsymbol{c})= \mathbb E[f(\boldsymbol{c})],
\end{equation}
where the expectation value is taken with respect to the distribution of data.  Naturally, due to finite sample statistics, this function is \emph{stochastic} and thus the optimization term ``function call'' takes on new meaning---the same set of controls $\boldsymbol{c}$ might result in different function evaluations $\hat f$.  Thus, the problem of maximizing the fidelity becomes one of \emph{stochastic optimization} \cite{spall2005introduction}.

\subsection{\emph{in situ} control algorithm}

A particularly useful set of techniques for multidimensional analysis goes by the name \emph{simultaneous perturbation stochastic approximation} (SPSA) \cite{spall1992multivariate}.  In short, SPSA is an iterative optimization technique which uses only two (noisy) function calls per iteration to estimate the gradient.  Our algorithm, ACRONYM, is direct quantum analog of SPSA.   The steps of each iteration are outlined as follows.  First fix a tolerance $\epsilon>0$.  For each $k$, so long as $\|\boldsymbol{c}_{k+1}-\boldsymbol{c}_{k}\|> \epsilon$, repeat:
\begin{enumerate}
\item Generate a random direction to search in defined by the vector $\boldsymbol{\triangle}_k$.  A recommend vector which we choose is where each element is selected according to a fair coin toss: $\boldsymbol{\triangle}_{kj} = \pm1$.
\item Calculate the estimated gradient 
\begin{equation}
\boldsymbol{g}_k = \frac{\hat f(\boldsymbol{c}_k+\beta_k\boldsymbol{\triangle}_k)-\hat f(\boldsymbol{c}_k-\beta_k\boldsymbol{\triangle}_k)}{2\beta_k}\boldsymbol{\triangle}_k.
\end{equation}
\item Calculate the next iterate via
\begin{equation}
\boldsymbol{c}_{k+1} = \boldsymbol{c}_{k} + \alpha_k \boldsymbol{g}_k. 
 \end{equation}
\end{enumerate}
The functions $\alpha_k$ and $\beta_k$ control the convergence and are user defined, although they are usually specified in the forms
\begin{equation}
\alpha_k = \frac{a}{(k+1)^s},\;\; \beta_k= \frac{b}{(k+1)^t},
\end{equation}
where $a,b,s$ and $t$ are chosen first roughly based on extensive numerical studies for many problems then tweaked based on numerical simulations for the problem at hand.  For reference, generally good choices are \cite{sadegh1996optimal} $a=b=1$, $s=0.602$ and $t=0.101$, but we have found that the asymptotically optimal values \cite{spall1992multivariate} $s=1$ and $t=1/6$ give good results across all parameter regimes considered here.     

\subsection{Convergence discussion}

There are a number of convergence results on stochastic optimization and the variant we use \cite{spall1992multivariate,sadegh1996optimal}.  All conclude that the error in the design space decreases at rate $O(k^{\beta_c})$ where typically $\beta_c\in[-1/2,0]$.  { Note that the lower bound comes from standard statistical arguments.}
The actual performance achieved depends on a number of often competing factors, thus we will allow the data to decide and compare to what might be expected from asymptotic arguments.

Here, we will be interested in the convergence in the objective function---the fidelity---rather than the controls.  { First we argue that $f$ will decrease as $O(k^{\beta_f})$ for $\beta_f = 2\beta_c$.  Assuming $f$ is differentiable and obtains its minimum at $\vec{c}_{\rm opt}$, the gradient $\nabla f (\vec{c}_{\rm opt}) = 0$ and we obtain the bound
\begin{align}
|f(\vec{c}_{k}) - f(\vec{c}_{\rm opt})| &\leq \frac{K}{2} \|H[ f](\vec{c}_{\rm opt}) \|  \|\vec{c}_{k}-\vec{c}_{\rm opt}\|_2^2,\\
& \frac{K}2 \|H[ f](\vec{c}_{\rm opt}) \| k^{2\beta_c},\label{eq:bound}
\end{align}
for some constant $K$ and where $H [f](\vec{c}_{\rm opt})$ is the Hessian of second derivatives, evaluated at $\vec{c}_{\rm opt}$ and $\|H[f](\vec{c}_{\rm opt})\|$ is the spectral norm, the largest eigenvalue, of $H[f](\vec{c}_{\rm opt})$.  Thus, we should still expect $O(k^{\beta_f})$ convergence in fidelity, with optimality given by $\beta_f\approx -1$.}  In the examples we consider, we will extract the exponents from fits to the simulation data.  Since we will not generally consider $\beta_c$, we drop the subscript $f$ on $\beta_f$ from now on.

{In discussing the performance of the algorithms considered here, we will refer to three numbers: $N$, the number of experiments used for each fidelity estimate; $M$, the number of fidelity estimates required for each iteration of the algorithm; and $k$, the number of iterations.  Thus, the total number of physical experiments required after $k$ iterations is $N_{\rm tot} = N\cdot M \cdot k$.  For ACRONYM, $M = 2$ regardless of the dimension of the control space or quantum system.  Moreover, we will show that the performance is roughly independent of $N$.  Thus we can limit our attention to the performance as a function of $k$.  For NM, the situation is more complicated since the performance depends crucially on $N$, and $M$ randomly fluctuates.  In all cases considered below, however, ACRONYM outperforms NM by any metric.}

\subsection{A tale of two fidelities}
{ 
We acknowledge the apparent contradiction in claiming that the fidelity converges when it is only known up to statistical fluctuations of order $1/\sqrt{N}$.  However, our claim is that the \emph{underlying} fidelity \eqref{eq:underlying} achieved by ACRONYM converges and not the \emph{estimated} fidelity \eqref{f+noise}---the estimated fidelity from noisy observation does not converge.  The latter situation is simply a consequence of finite statistics \emph{per iteration}.  

Though both fidelities are important in their own right, the problem of control \emph{is} to maximize the underlying fidelity, rather than to verify it.  As a consequence, the fidelity of the currently selected control may be much closer to the optimal fidelity than can be verified by experiments with the same noise sources.  In an experiment, the experimenter must trust that the algorithm converges or verify the controls found at the final iteration with more resources than used in their finding.  In settings where quantum control is expected to be an automated subroutine, verifying the solution would not be expected in production.
}

\section{Numerical experiments\label{sec:examples}}

\begin{figure*}[ht]
\includegraphics[width=\linewidth]{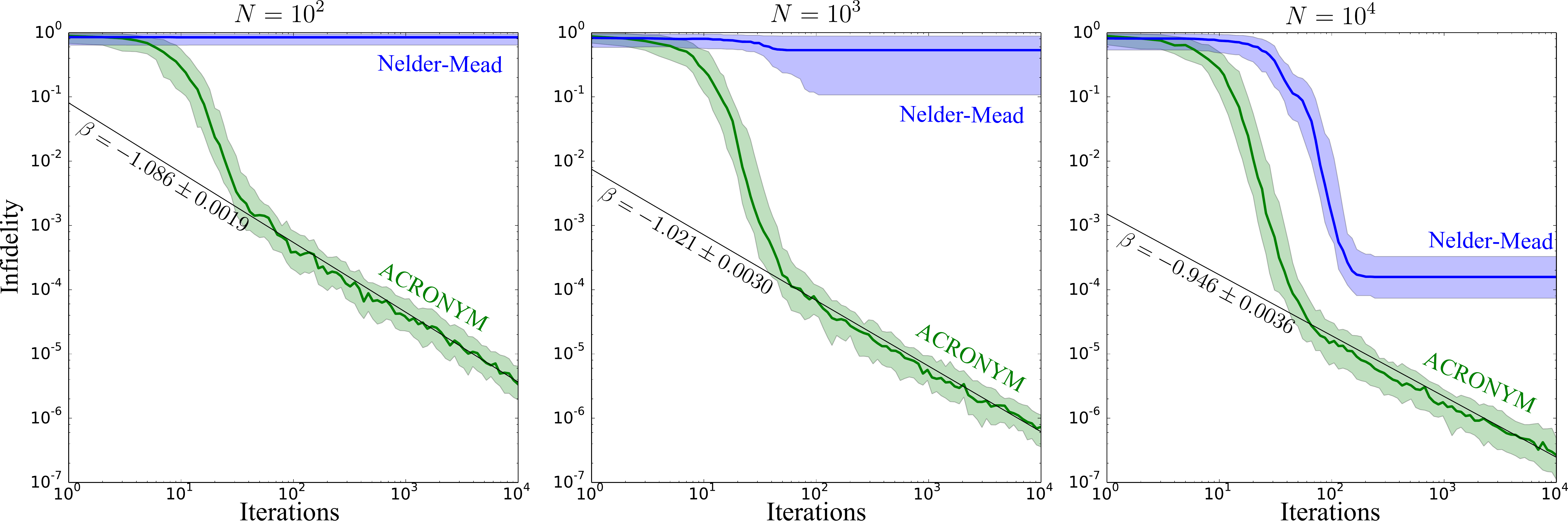}
\caption{\label{fig:1qubit_finiteN} (Color online)
For the single qubit random gate design problem, plotted is the \emph{infidelity} (1-fidelity) of the ACRONYM and NM algorithms as a function of the number of iterations for three different numbers of experiments per iteration: (Left to right) $10^2$, $10^3$ and $10^4$.  The thick lines are the median of the data and the shaded region around each is the interquartile range (the middle 50\% of the data).  The thin black lines are fits to the theoretical $O(k^\beta)$ scaling.  We see that NM is limited by the noise in the function call while ACRONYM is not.}  
\end{figure*}

\begin{figure}[ht]
\includegraphics[width=.7\columnwidth]{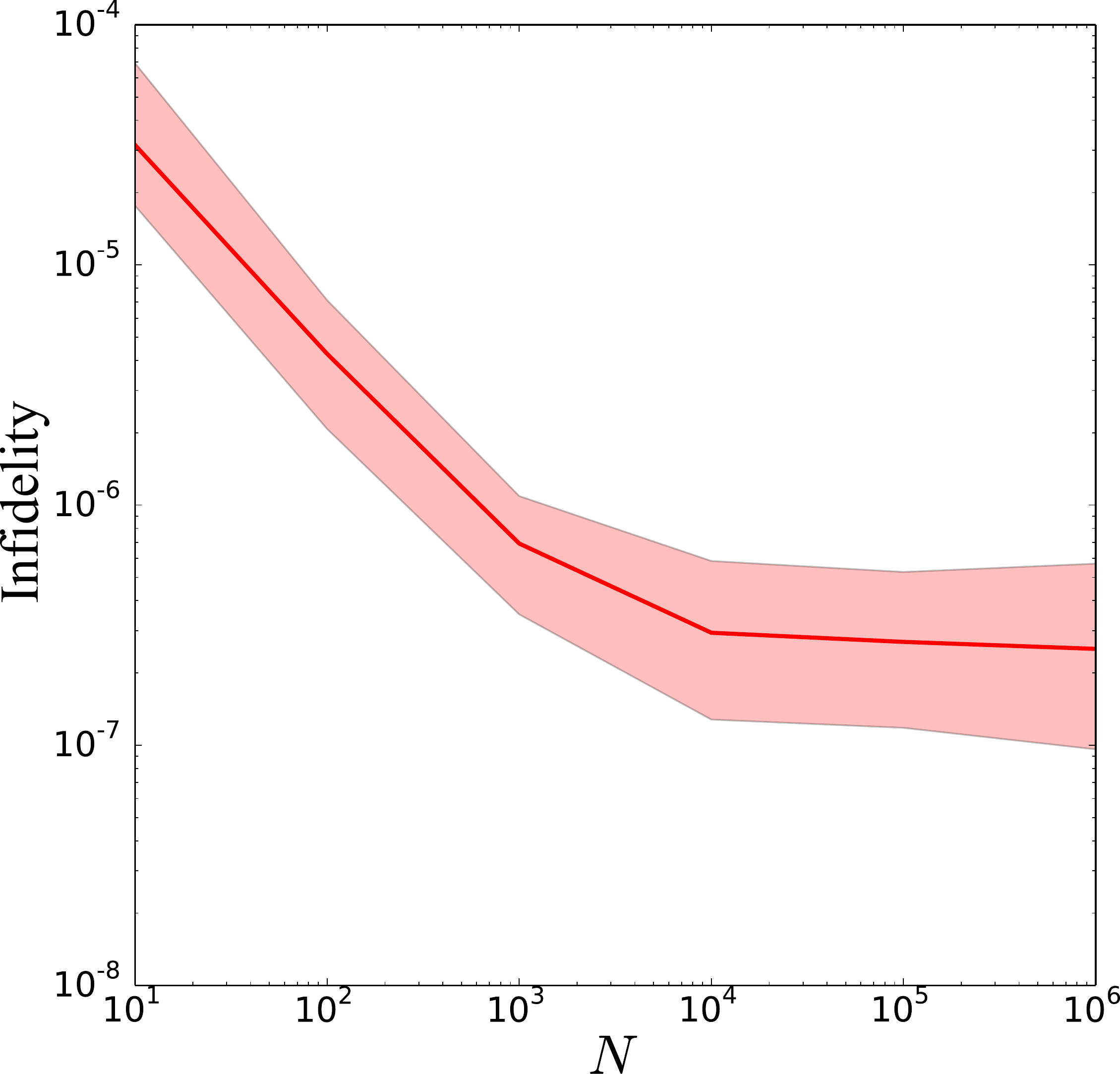}
\caption{\label{fig:qubit_v_N} (Color online)
For the single qubit random gate design problem, plotted is the \emph{infidelity} (1-fidelity) of the ACRONYM as a function of $N$, the number of experiments per fidelity estimate at $k = 10^4$ iterations. The dimension of the control space is $p = 10$.  The thick line is the median of the data and the shaded region is the interquartile range (the middle 50\% of the data). This demonstrates, with the results of Fig. \ref{fig:1qubit_finiteN}, that more experiments per iteration of the algorithm are not necessary to converge.  At some point (here $\approx 10^4$), a constant number of experiments suffices.}
\end{figure}

\begin{figure*}[ht]
\includegraphics[width=1.5\columnwidth]{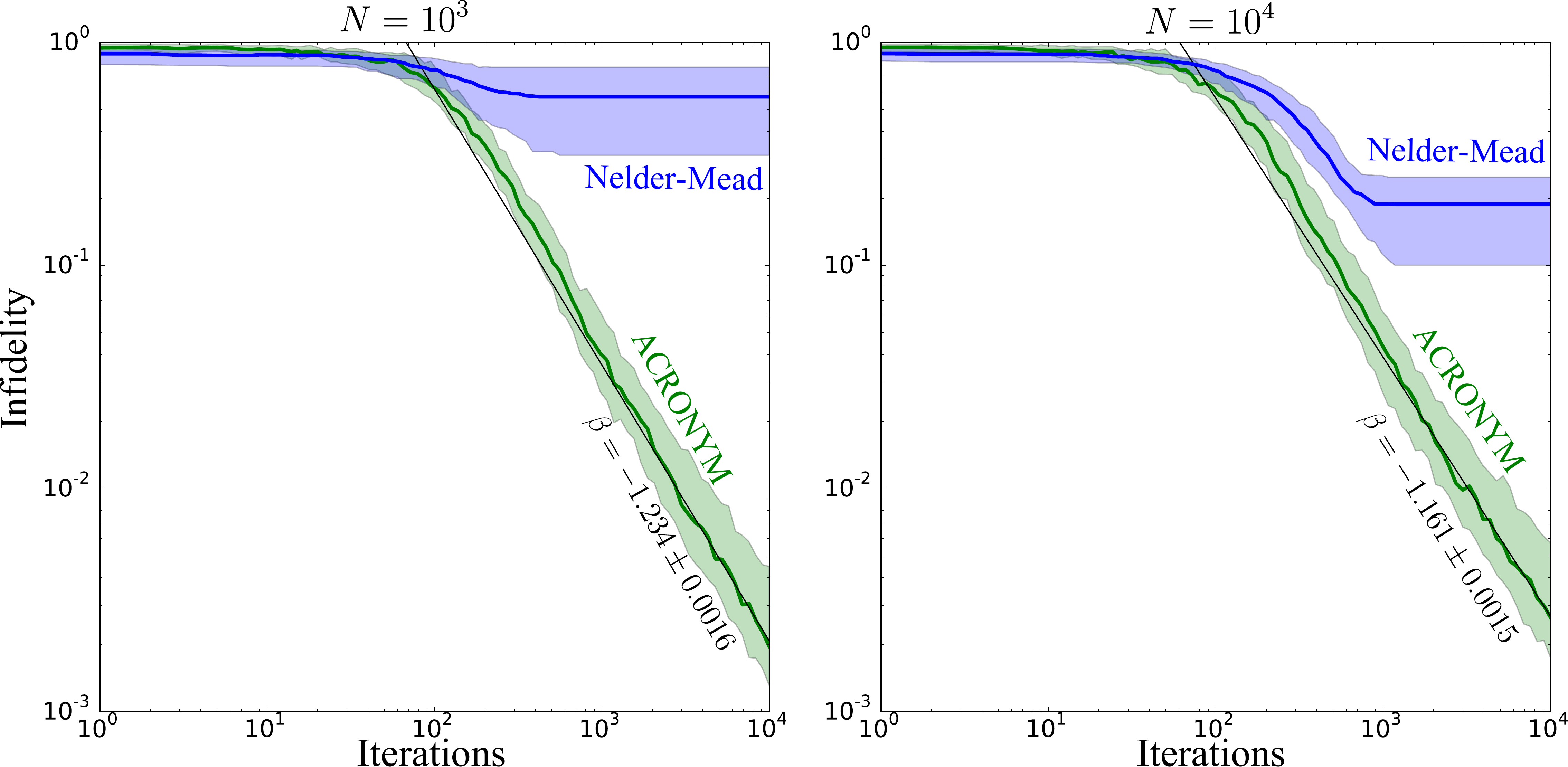}
\caption{\label{fig:2qubit_CNOT} (Color online)
For the control problem of designing a C-NOT gate, plotted is the \emph{infidelity} (1-fidelity) of ACRONYM and NM as a function of the number of iterations for $N= 10^3$ and $N=10^4$ experiments per iteration.  The thick lines are the median of the data and the shaded region is the interquartile range (the middle 50\% of the data).  As shown for the single qubit problem in the figures above, ACRONYM converges while NM does not.}
\end{figure*}

\subsection{Random single-qubit gates}

The simplest demonstration is that of a qubit subjected to a Hamiltonian, in some frame,
\begin{equation}
H(t) = \omega_0 Z + \omega_c(t) X,
\end{equation}
where $Z$ and $X$ are the single qubit Pauli operators, $\omega_0 Z$ is the fixed drift Hamiltonian, and $\omega_c (t) X$  is the control.  The task is to choose $\omega_c$ such that
\begin{equation}
U_c = \exp(-iH(t)) = U_{\rm T}.
\end{equation}
Even for this simple problem, there is no closed-form solution, which is why we consider problem of fidelity optimization.  For this example, we will average over targets randomly generated from the Haar measure.  We work in dimensionless units where $\omega_0 = 1$ and the control is piecewise constant~\cite{note_parametrization} on the time intervals between $\{0,1,2,\ldots,p\}$ so that $\vec{c} = (\omega_c(1), \omega_c(2),\ldots, \omega_c(p))\in \mathbb R^p$.  We assume the fidelity is estimated via finite sample statistics such that the objective function in Eq.~\eqref{f+noise} is distributed according to $N \hat f(\vec{c}) \sim {\rm Bin}(N,f(\vec{c}))$, a Binomial distribution, where $N$ is the number of experiments used to estimate the fidelity and the Binomial parameter is the true fidelity of the controls $\vec{c}$.  {This model for fidelity estimation was chosen as it conveniently and intuitively demonstrates the key important limitation of using non-stochastic optimization techniques: that fidelity must be estimated from experimental samples.  The actual model will vary depending on the details of the experimental implementation, but what remains is that fidelity estimation is limited by statistical fluctuations---and here we will explore the ultimate shot-noise limits.}

First, we demonstrate the utility of \emph{in situ} control in general before comparing different \emph{in situ} strategies.  Suppose in an \emph{ex situ} algorithm (such as \cite{khaneja2005optimal}), the model above is not quite correct.  For example, suppose that the drift Hamiltonian is incorrectly modeled such that $H_{\rm drift} = Z + \triangle H$, for some random perturbation $\triangle H$.  In particular, suppose $\|\triangle H\| = 0.01$.  Then the offline optimization will rapidly design a control scheme for this model to hit the target $U_{\rm T}$.  However, these controls will implement a different unitary under the dynamics of the true model.  On the other hand, ACRONYM uses the true model by fiat, but, as any \emph{in situ} protocol, suffers a penalty for not being able to exactly compute the objective function---it must do so through experimental trials.  In Fig.~\ref{fig:first}, we show the performance of an offline algorithm with a bad model and ACRONYM.  As expected, the offline algorithm does quite well until the errors in the model dominate.  On the other hand, ACRONYM continues to learn the optimal controls.  Next, we will determine the asymptotic rate of learning and compare ACRONYM to the NM algorithm.

In the following, then, we report on numerical experiments which demonstrate that ACRONYM converges in fidelity to the target at rate $O(k^\beta)$ given by $\beta \approx - 1$.  Since we show the asymptotic performance is independent of the the number of experiments per iteration, the overall performance appears to scale as $O( 1/N_{\rm tot})$. The simulations and our optimization algorithm, as well as comparisons~\cite{note_neldermead} to the NM algorithm, were implemented in Python using SciPy \cite{scipy}.  The results for the single qubit problem are summarized in Fig.~\ref{fig:1qubit_finiteN}.  As expected, the more experimental samples taken per iteration, the less the statistical noise, and the higher the fidelity given a fixed number of iterations for NM.  We also see that NM optimization ceases to improve when the statistical noise dominates its ability to guess the correct search direction.

For all qubit simulations the control space dimension is $p = 10$ which would naively suggest that $M= 20$ fidelity evaluations are required to estimate the gradient.  We will discuss the NM algorithm further in the discussion of Section \ref{sec:end}, but here we note that for the simulations presented in Fig.~\ref{fig:1qubit_finiteN} the average number of function calls per iteration used by NM was $M =3.13$, 50\% more than ACRONYM which, recall, requires exactly two fidelity estimates, $M = 2$.  Moreover, ACRONYM's performance is roughly independent of the number of experiments per iteration, as shown in Fig.~\ref{fig:qubit_v_N}.  We see that after $N \approx 10^4$ experiments per fidelity estimate, there is no additional gain in performance.  This is important as it implies that very few overall experiments indeed are needed to converge to the target unitary.  This is another way of seeing that additional experiments are not needed at each iteration in order to converge---a constant number of experiments suffices to converge.

\subsection{C-NOT gate}

In the first example, we considered single qubit gates, which are not sufficient for quantum computation.  To enable universal quantum computing we require a two-qubit entangling gate, the most commonly considered of which is the C-NOT \cite{nielsen2010quantum}.  Let the target $U_{\rm T}$ then be the C-NOT gate and the Hamiltonian be
\begin{align}
H(t) &= H_{\rm drift} + H_{\vec{c}}(t),
\end{align}
where the drift Hamiltonian consists of local Zeeman terms and a Heisenberg exchange interaction:
\begin{equation}
H_{\rm drift} = \delta_1 Z_1 + \delta_2  Z_2 + \frac J2 (X_1 X_2 + Y_1  Y_2 + Z_2  Z_2),
\end{equation}
and the control contains the transverse terms
\begin{equation}
H_{\vec{c}} = c_{x,1}(t) X_1 + c_{x,2}(t)  X_2 + c_{y,1}(t) Y_1 + c_{y,2}(t)  Y_2,
\end{equation}
where the subscripts denote which qubit the operator acts on.  For the simulation, we take $\delta_1 = -1\delta_2 = 10 J =1$ and have the controls be piecewise constant on the intervals $\{0,1,2,\ldots,q\}$ so that
\begin{align}
\vec{c} = (&c_{x,1}(1), c_{x,1}(2),\ldots, c_{x,1}(q),\\
&c_{x,2}(1),\ldots, c_{y,1}(1), \ldots, c_{y,2}(1),\ldots, c_{y,2}(q)),\nonumber
\end{align}
which is now a vector in $\mathbb R^p$ with $p=4q$.

Again, we consider the case of estimating the fidelity from finitely many experiments.  In Fig.~\ref{fig:2qubit_CNOT}, the infidelity is plotted versus the number of iterations for both  $N = 10^3$ and $N = 10^4$ number of experiments per fidelity for both ACRONYM and NM.  In this case we used $q=10$, which means the dimension of the search space was $p  = 40$.  A standard finite difference gradient approximation would require $M = 80$ fidelity estimations per iteration while ACRONYM requires exactly $M = 2$ (for this problem NM used an average of $M = 3.23$ function calls per iteration---62\% more than ACRONYM).  From Fig.~\ref{fig:2qubit_CNOT}, we see that ACRONYM continues to convergence after NM has saturated which occurs quite rapidly for even $N = 10^4$ experiments per fidelity estimate.  The convergence of ACRONYM is again $O(k^\beta)$ with fits giving $\beta \approx -1$.  { These numerical examples suggest
ACRONYM has superior performance for \emph{in situ} control.
}
\subsection{Robustness to control noise}

\begin{figure}[h]
\includegraphics[width=.75\columnwidth]{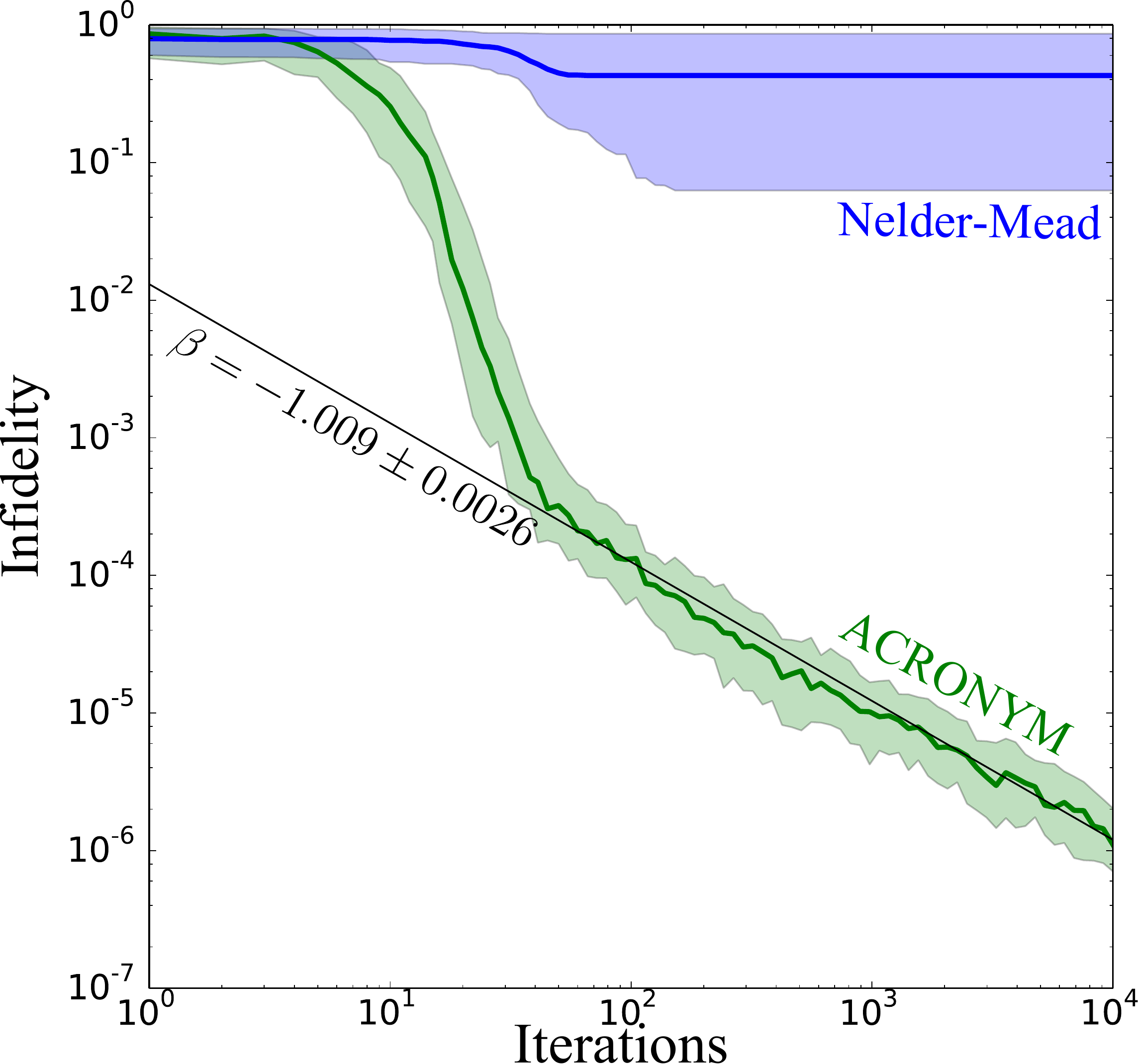}
\caption{\label{fig:1qubit_noisy_controls} (Color online)
Back to the single qubit problem, now with added with control noise.  Plotted is the \emph{infidelity} (1-fidelity) of the ACRONYM and NM algorithms as a function of the number of iterations.  ACRONYM used $N = 10^3$ experiments per function evaluation, but NM was given exact fidelity values.  Both methods were subjected to independent zero-mean Gaussian noise on each control component with standard deviation $10^{-2}$.  Even with exactly fidelity values, NM is limited by the control noise.  ACRONYM is robust to both statistical and control noise.}
\end{figure}

Finally, we show that ACRONYM is robust by considering an additional source of noise, namely imperfect implementation of the controls.  To model this, we add to every experimental iteration independent zero-mean Gaussian noise on each component $\vec{c}_j$ of the control vector.  { To be precise, each fidelity estimate is expected to be performed using the control parameters $\vec{c}$, but what actually happens is $\vec{c} +\vec{\epsilon}$, where $\vec{\epsilon}$ is drawn according to a zero-mean multivariate normal distribution with diagonal covariances of $10^{-2}$.}  The results are shown in Fig.~\ref{fig:1qubit_noisy_controls}.  As in Fig.~\ref{fig:1qubit_finiteN} (right), ACRONYM used $N = 10^4$ experiments for each fidelity evaluation (2 per iteration, recall).  However, we gave NM exact fidelity values---essentially infinitely many experiments for free.  Both methods were then subjected to control noise of strength (standard deviation) $10^{-2}$.  We see that even when NM has perfect fidelity evaluation, the added noise renders the method useless (additional noise on fidelity estimation leave the NM approach learning nothing at all).  Whereas, ACRONYM is robust to the added noise maintaining converge at a rate given by $\beta \approx -1$.  This shows that ACRONYM is robust to many simultaneous sources of noise, both statistical and technical. 

\section{Conclusion and discussion\label{sec:end}}

Before we conclude, we distinguish some further discussion points, speculation and directions for future research.

\subsection{More on noise}

So far, we have assumed that the noise on the fidelity estimation instances is completely independent and identically distributed, and we have shown that ACRONYM, by virtue of accounting for the stochasticity of the fidelity evaluations, is not limited by this type of noise. However, in real experiments, noise comes in more flavors.  So, what other kinds of noise is ACRONYM robust to?  The original convergence proof of Spall \cite{spall1992multivariate} only required that the expected difference of the noise at each function evaluation be zero.  In other words, the noise need not assumed independent.  Thus, for example, low-frequency drifts should not be a problem, but higher-frequency drifts may result in a bias in the gradient estimate at each evaluation.  Conveniently, in most scenarios, it is the low frequency noise that is more difficult to detect and correct while the problem high frequency noise is more routinely dealt with.  

{
Another common example of noise in the quantum device characterization literature is state preparation and measurement (SPAM) error.  As demonstrated in randomized benchmarking and \emph{self-consistent} approaches \cite{gambetta,rbk}, however, SPAM errors can be cleverly mitigated.  In any case, the story remains the same as above: if the noise biases the gradient estimates, the algorithm will not necessarily converge.
}

{  We note that while ACRONYM solves the optimization problem even in the presence of additional control noise, the final implementation itself will not be robust to the same noise.  That is, ACRONYM finds the controls which would have the highest fidelity without noise, but does not find a set of controls have the same robustness properties as typically sought after, such as optimal average or worst-case fidelity when varying over the distribution of noise (see, for example, \cite{noise1,noise2}).  In a sense, ACRONYM solves the problem---in the presence of noise---\emph{as if} the noise were not present.  One can imagine, for example, using such a technique to ``tune-up'' gates first in the presence of noise before spending resources to fine-tune the control mechanism.  On the other hand, it would also be interesting to combine this approach to the robustness techniques in, for example, \cite{noise1,noise2}.}

\subsection{More on Nelder-Mead}

While ACRONYM requires only $M=2$ function calls per iteration, NM is still doing impressively well at only $M\approx 3.5$, on average.  Unfortunately, there is little more we can say since, although NM is widely popular and successful, it still lacks (after 50 years!) a satisfactory convergence proof \cite{nm}.  Moreover, in many applications, it is found to underperform in high dimensions \cite{nm}.  

As countless others have empirically observed, NM does work very well in some cases.  In fact, we have seen that there are some regimes where NM outperforms ACRONYM.  These are typically earlier in the search where the noise in the function call is dominated by the distance in the objective function to the optimal point.  Perhaps, then, a more efficient adaptive protocol exists which begins with NM and switches to ACRONYM?

Although we have harped on NM for not performing well in the presence of noise, we do note that some variants of NM have been proposed to deal with stochasticity in the objective function \cite{snm}.  It is unclear how well such methods would perform on quantum control problems. 

{
\subsection{Learning-type control}

Recall the update rule for the controls:
\begin{equation}
\boldsymbol{c}_{k+1} = \boldsymbol{c}_{k} + \alpha_k \boldsymbol{g}_k.
\end{equation}
Generically, one could have a law of the form:
\begin{equation}
\boldsymbol{c}_{k+1} = \mathcal A(\boldsymbol{c}_{k}),
\end{equation}
with the obvious demand that the optimal solution be the unique fixed point of $\mathcal A$.  Specific algorithms in classical control theory of this type, dubbed \emph{learning-type control}, typically use linear models for the output (see, for example, \cite{r2r}) and are hence not immediately applicable to the quantum control problem.  It would be interesting, however, to look to this more general formalism for potential improvements over ACRONYM.  Ideally, we would like to replace the two function calls demanded by SPSA by one---perhaps allowing that the updated controls are a specification of experiment to be performed.  This would further minimize the role of additional classical computation and may give new insights into quantum automation and learning problems.
}

\subsection{Parametrization freedom\label{subsec:freedom}}

The \emph{ex situ} paradigm of pulse design, due to the enduring inefficiency of simulating arbitrary time-dependent Hamiltonians on a classical computer, imposes a preference on the parametrization of the controls. In most cases, the preferred parametrization is one that leads to the Hamiltonian being piecewise time-independent in some frame, simplifying the computation of the overall quantum propagator on a classical computer (as a time-ordered product of exponentials instead of the more costly time-ordered exponentiation of an integral.) This parametrization is further encouraged by the ability to approximate the gradient of the fidelity given the already computed step-wise propagators~\cite{khaneja2005optimal}. 

Switching to an \emph{in situ} setting allows for more freedom in the choice of parametrization, for one is no longer required to simulate the quantum dynamics on a classical computer. That is to say, the choice of the parametrization can be driven by other requirements or preferences. For example, one could potentially choose a parametrization with few parameters; greatly reducing the parameter space over which the optimization is performed. Another could look for a parametrization that has a better fidelity landscape; making the identification or optimization problem easier. Yet another parametrization could more readily and naturally specify other constraints on the pulse form like frequency bandwidth or frequency selectivity. In general, we posit that any parametrization that deterministically maps the parameter space onto realizable wave forms can be used with \emph{in situ} algorithms.

It is understood of course that time-ordering is important in quantum mechanics. Yet, one wonders if thinking about global approaches as opposed to temporally-local parametrizations would be beneficial, since, after all, the goal is to design an overall propagator as opposed to an instantaneous Hamiltonian. That is to say, what matters in quantum gate design is the destination not the route. 

At any rate, we expect that this freedom of parametrization will allow researchers to choose parametrizations that are more natural or better suited to their specific problems, and will encourage the diversification of approaches, from which an evolved approach better suited to the general problem of pulse design will arise.

\subsection{Conclusion}
In this work, we have introduced ACRONYM, a stochastic optimization algorithm to design \emph{in situ} control sequences for quantum information processing tasks.  The fact that fidelity estimation can be done efficiently (via randomized benchmarking, for example) and our algorithm requires a constant number of experiments per iteration---regardless of the dimension of the control space---implies that ACRONYM is efficient.  We have also demonstrated that it is robust not only to the statistical noise inherent in \emph{in situ} fidelity estimation but also to noise on the control fields. Moreover, ACRONYM is implementation independent---it requires only that the controls produce changes in the fidelity which can be estimated via experiment.

\begin{acknowledgements}
O.M. acknowledges the difficulty inherent in pinpointing the time of birth of ideas, let alone the moment of conception of insight, and in this light, thanks everyone he has talked with over the years about \emph{in situ} control, in particular M.~P.~da~Silva, C.~A.~Ryan, J.~Emerson, R.~Laflamme, and D.~G.~Cory.  
CF was supported by National Science Foundation grant number PHY-1212445, the Canadian Government through the NSERC PDF program, the IARPA MQCO program, the ARC via EQuS project number CE11001013, and by the US Army Research Office grant numbers W911NF-14-1-0098 and W911NF-14-1-0103.
\end{acknowledgements}

\end{document}